\documentclass{aa}
\input{epsf}
\newcommand{\EQ}{\begin{equation}}
\newcommand{\EN}{\end{equation}}
\newcommand{\EQA}{\begin{eqnarray}}
\newcommand{\ENA}{\end{eqnarray}}
\newcommand{\eq}[1]{(\ref{#1})}
\newcommand{\Eq}[1]{Eq.~(\ref{#1})}
\newcommand{\Eqs}[2]{Eqs.~(\ref{#1}) and~(\ref{#2})}
\newcommand{\Sec}[1]{Sect.~\ref{#1}}
\newcommand{\Fig}[1]{Fig.~\ref{#1}}

\newcommand{\bra}[1]{\langle #1\rangle}

\newcommand{\uu}{\mbox{\boldmath $u$} {}}

\newcommand{\BB}{\mbox{\boldmath $B$} {}}

\newcommand{\AAA}{\mbox{\boldmath $A$} {}}

\newcommand{\JJ}{\mbox{\boldmath $J$} {}}

\newcommand{\ff}{\mbox{\boldmath $f$} {}}

\newcommand{\nab}{\mbox{\boldmath $\nabla$} {}}

\newcommand{\oo}{\mbox{\boldmath $\omega$} {}}

\newcommand{\DD}{{\rm D} \, {}}
\newcommand{\dd}{{\rm d} {}}

\def\la{\mathrel{\mathchoice {\vcenter{\offinterlineskip\halign{\hfil
$\displaystyle##$\hfil\cr<\cr\sim\cr}}}
{\vcenter{\offinterlineskip\halign{\hfil$\textstyle##$\hfil\cr<\cr\sim\cr}}}
{\vcenter{\offinterlineskip\halign{\hfil$\scriptstyle##$\hfil\cr<\cr\sim\cr}}}
{\vcenter{\offinterlineskip\halign{\hfil$\scriptscriptstyle##$\hfil\cr<\cr\sim\cr}}}}}

\newcommand{\ea}{{\rm et al. }}

\def\onethird{{\textstyle{1\over3}}}

\newcommand{\yapj}[3]{ #1, {ApJ }{#2}, #3}
\newcommand{\yapjl}[3]{ #1, {ApJ (Letters) }{#2}, #3}

\newcommand{\yana}[3]{ #1, {A\&A }{#2}, #3}

\newcommand{\yjfm}[3]{ #1, {J. Fluid Mech. }{#2}, #3}

\newcommand{\yjetp}[3]{ #1, {Sov. Phys. JETP }{#2}, #3}

\newcommand{\yprl}[3]{ #1, {Phys. Rev. Lett. }{#2}, #3}
\newcommand{\yphl}[3]{ #1, {Phys. Lett. }{#2}, #3}

\newcommand{\ynat}[3]{ #1, {Nat }{#2}, #3}

\newcommand{\ybook}[3]{ #1, {#2} (#3)}

\begin{document}
\title{Large scale dynamos with ambipolar diffusion nonlinearity}
\author{Axel Brandenburg\inst{1,2} and Kandaswamy Subramanian\inst{3}}
\authorrunning{A.\ Brandenburg and K.\ Subramanian}
\institute{
NORDITA, Blegdamsvej 17, DK-2100 Copenhagen \O, Denmark
\and Department of Mathematics, University of Newcastle upon Tyne, NE1 7RU, UK
\and National Centre for Radio Astrophysics - TIFR, Poona University Campus,
Ganeshkhind, Pune 411 007, India
}

\maketitle

\begin{abstract}
It is shown that ambipolar diffusion as a toy nonlinearity leads to
very similar behaviour of large scale turbulent dynamos as full MHD.
This is demonstrated using both direct simulations in a periodic box
and a closure model for the magnetic correlation functions applicable
to infinite space. Large scale fields develop via a nonlocal inverse
cascade as described by the alpha-effect. However, because magnetic
helicity can only change on a resistive timescale, the time it takes
to organize the field into large scales increases with magnetic
Reynolds number.
\end{abstract}

\section{Ambipolar diffusion as a toy nonlinearity}

In this Letter we test and exploit the idea that the exact type of
nonlinearity in the MHD equations is unessential as far as the nature
of large scale field generation is concerned. At first glance this may
seem rather surprising, especially if one pictures large scale field
generation as the result of an inverse cascade process (Frisch \ea 1975,
Pouquet \ea 1976). Like the direct cascade in Kolmogorov turbulence,
the inverse cascade is accomplished by nonlinear interactions, suggesting
that nonlinearity is important. However,
a special type of inverse cascade is the strongly nonlocal inverse
cascade process, which is usually referred to as the `alpha-effect';
see Moffatt (1978) and Krause \& R\"adler (1980). This effect exists
already in linear (kinematic) theory.

Until recently it was unclear which, if any, of the two effects
(inverse cascade in the local sense or the $\alpha$-effect) played
the dominant role in large scale field generation as seen in simulations
(e.g.\ Glatzmaier \& Roberts 1995, Brandenburg \ea 1995, Ziegler \&
R\"udiger 2000) or in astrophysical bodies (stars, galaxies, accretion
discs). A strong indication that it is actually the $\alpha$-effect
(i.e.\ the strongly nonlocal inverse cascade) that is responsible for
large scale field generation, comes from detailed analysis of recent
three-dimensional simulations of forced isotropic non-mirror symmetric
turbulence (Brandenburg 2000, hereafter B2000). In those
simulations a strong and nearly force-free magnetic field was produced, 
and most of the energy supply to this field
was found to come from the forcing scale of the turbulence.

In the absence of nonlinearity, however, the field seen in the simulations
of B2000 became quickly swamped by magnetic fields at
smaller scales. In that sense {\it a purely kinematic
large scale turbulent dynamo is impossible}!
Any hope for analytic progress is therefore slim. However, the
model of Subramanian (1997, 1999) is an exception. 
Subramanian (1997; hereafter S97) extended the kinematic models of
of Kazantsev (1968) and Vainshtein \& Kitchatinov (1986) by including ambipolar 
diffusion (in the strong coupling approximation) as a nonlinearity. Under
the common assumption that the velocity is delta-correlated in time,
S97 derived a nonlinear equation for the
evolution of the correlation functions of magnetic field and magnetic
helicity. Although the models of Kazantsev (1968) and Novikov \ea (1983)
are usually known to describe small-scale field generation, Subramanian
(1999; hereafter S99) found that in the presence of fluid helicity there 
is the possibility of tunnelling of bound-states corresponding to small 
scales to unbounded states corresponding to large scale fields,
which are force-free.

In this Letter we present numerical solutions to the closure model of
S99. We stress that we do {\it not} advocate ambipolar diffusion (AD)
as being dominant over the usual feedback from the Lorentz force in the
momentum equation. Instead, our motivation is to establish a useful {\it
toy model} to study effects of nonlinearity in dynamos. Our
numerical solutions may provide guidance for further analytic treatment
of these equations in parameter regimes otherwise inaccessible. We begin
however by considering first solutions of the fully three-dimensional
MHD equations in a periodic box using AD as the only nonlinearity.

\section{Box simulations for a finite system}
\label{Speribox}

In this section we adopt the MHD equations for an isothermal compressible
gas, driven by a given body force $\ff$, in the presence of AD,
but ignoring the Lorentz force
\EQ
{\DD\ln\rho\over\DD t}=-\nab\cdot\uu,
\EN
\EQ
{\DD\uu\over\DD t}=-c_{\rm s}^2\nab\ln\rho
+{\mu\over\rho}(\nabla^2\uu+\onethird\nab\nab\cdot\uu)+\ff,
\label{dudt}
\EN
\EQ
{\partial\AAA\over\partial t}=(\uu+\uu_{\rm D})\times\BB
-\eta\mu_0\JJ,
\label{dAdt}
\EN
where ${\rm D}/{\rm D}t=\partial/\partial t+\uu\cdot\nab$ is the
advective derivative, $\BB=\nab\times\AAA$ is the magnetic field,
$\JJ=\nab\times\BB/\mu_0$ is the current density, and $\ff$ is the
random forcing function as specified in B2000. The nonlinear drift
velocity $\uu_{\rm D}$ due to AD can be written as
$\uu_{\rm D}=a\,\JJ\times\BB$.
We use nondimensional units where $c_{\rm s}=k_1=\rho_0=\mu_0=1$.
Here, $c_{\rm s}$ is the sound speed, $k_1$ the smallest wavenumber
of the box (so its size is $2\pi$), $\rho_0$ is the mean density,
and $\mu_0$ is the vacuum permeability. Since AD is the only
nonlinearity in \Eq{dAdt} we can always normalize $\BB$ such that
$a=1$.

\begin{figure}[h!]\epsfbox{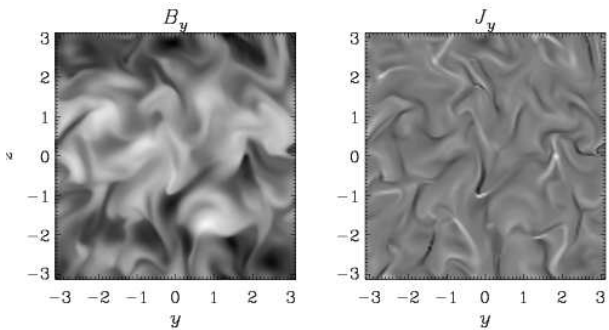}\caption[]{
Images of $B_y$ and $J_y$ in an arbitrarily chosen $yz$ plane.
$120^3$ meshpoints, $t=337$.
}\label{Fbfield}\end{figure}

\begin{figure}[t]\epsfbox{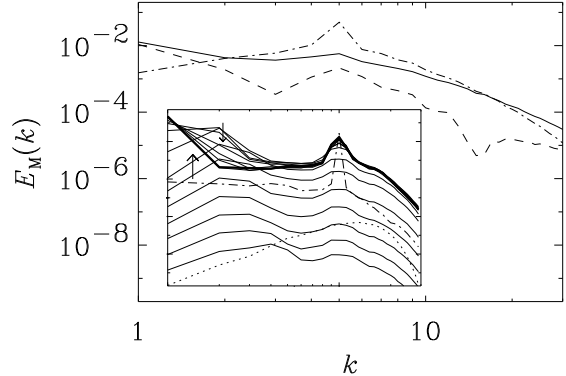}\caption[]{
Spectra of magnetic energy (solid lines), kinetic energy (dash-dotted line),
magnetic helicity (normalized by $k/2$; dashed line) for the run shown in
\Fig{Fbfield}. The inset shows spectra of a run with forcing at $k=10$ and
$60^3$ meshpoints for different times till $t=900$.
}\label{Fpp}\end{figure}

The model presented here is similar to Run~3 of B2000, where $\mu =
\eta=2\times10^{-3}$. With a root-mean-square velocity of around 0.3 the
magnetic Reynolds number based on the size of the box is around 1000. The
forcing wavenumber $k_{\rm f}$ is chosen to be 5. In \Fig{Fbfield}
we show a grey scale representation of a slice of the magnetic field
and the current density at $t=337$. Note the presence of a large scale
magnetic field that varies in the $z$-direction. In \Fig{Fpp} we show
the spectra of magnetic and kinetic energies. The peak of magnetic energy
at $k=1$ shows the development of large scale magnetic fields. Further,
the current density is concentrated into narrow filamentary structures,
typical of AD (see Brandenburg \& Zweibel 1994).

\epsfxsize=8.2cm

Unfortunately, the severity of the diffusive timestep limit, $\delta t\le
0.16\delta x^2/\eta_{\rm AD}$, where $\eta_{\rm AD}=a\BB^2$, prevented
us from running much longer at high resolution ($120^3$ meshpoints). For
$60^3$ meshpoints this limit is unimportant, and so we were able to run
until $t=900$, a time when the large scale field was much more clearly
defined. In the inset of Fig. 2, we show the evolution for such a case, 
but with a forcing at $k_{\rm f}=10$ (giving larger scale separation). 
Note again the peak of $E_M$ at $k=1$ and also the suppression of magnetic 
field at the next smaller scale, corresponding to $k\ge2$. Both these 
features are very similar to the magnetic field evolution in the case 
with full Lorentz force and without AD (Figs~3 and 17 of B2000).

\epsfxsize=8.4cm

Our main conclusion from these results is first of all that
large scale field generation works in spite of AD, contrary
to earlier suggestions that AD might suppress the large scale 
dynamo process (Kulsrud \& Anderson 1992). Secondly, AD provides 
a nonlinear saturation mechanism for the magnetic field at all scales, 
except for the scale of the box, where a force-free field develops 
for which $\uu_{\rm D}$ vanishes. Like in the simulations of B2000 this 
provides a `self-cleaning' mechanism, without which the field would 
be dominated by contributions from small scales.

Having established the close similarity between models with AD versus full
Lorentz force as nonlinearity, we now move on to discuss the nonlinear
closure model of S99 with AD as a `toy' nonlinearity.

\section{Closure model for an infinite system}

Under the assumptions that the velocity is delta-correlated in time
and the magnetic field is a gaussian random field S97 derived
equations for the longitudinal correlation function $M(r,t)$ and the
correlation function for magnetic helicity density, $N(r,t)$. The
velocity is represented by a longitudinal correlation function $T(r)$
and a correlation function for the kinetic helicity density, $C(r)$. We
change somewhat the notation of S99 and define the operators
\EQ
\tilde{D}(\cdot)={1\over r^4}\,{\partial\over\partial r}\left(r^4\cdot\right),
\quad D(\cdot)={\partial\over\partial r}(\cdot),
\EN
so the closure equations can be written as
\EQ
\dot{M}=2\tilde{D}(\eta_{\rm T}DM)+2GM+4\alpha H,
\label{closure1}
\EN
\EQ
\dot{N}=-2\eta_{\rm T} H+\alpha M,
\label{closure2}
\EN
where $H = - \tilde{D}DN$ is the correlation function of the current
helicity, $G=-\tilde{D}DT$ is the effective induction,
\EQ
\alpha=\alpha_0(r) + 4aH(0,t) 
\EN
\EQ
\eta_{\rm T}=\eta+ \eta_0(r) + 2 a M(0,t) 
\EN
are functions resembling the usual $\alpha$-effect and
the total magnetic diffusivity. Here $\alpha_0(r)=-2[C(0)-C(r)]$ and
$\eta_0(r)=T(0)-T(r)$. Note that at large scales
\EQ
\alpha_\infty\equiv\alpha(r\rightarrow\infty)=
-{\textstyle{1\over3}}\tau\bra{\oo\cdot\uu}
+{\textstyle{1\over3}}\tau_{\rm AD}\bra{\JJ\cdot\BB}/\rho_0
\label{asuppress}
\EN
\EQ
\eta_\infty\equiv\eta_{\rm T}(r\rightarrow\infty)=
{\textstyle{1\over3}}\tau\bra{\uu^2}
+{\textstyle{1\over3}}\tau_{\rm AD}\bra{\BB^2}/\mu_0\rho_0,
\EN
where $\tau_{\rm AD}=2a\rho_0$. Expression \eq{asuppress} is very
similar to the $\alpha$-suppression formula first found by
Pouquet \ea (1976). Here $\alpha$ and $\eta_{\rm T}$ are scale 
dependent (i.e.\ they are largest on large scales) and, in addition, 
both are affected by AD.

We construct $T(r)$ and $C(r)$ from an analytic approximation of the
kinetic energy and helicity spectra, $E_{\rm K}(k)$ and $H_{\rm K}(k)$,
respectively.  Zero velocity at large scales means that $E_K(k)\sim k^4$
for $k\rightarrow0$. At some wavenumber $k=k_{\rm f}$ the spectrum turns
to a $k^{-5/3}$ Kolmogorov spectrum, followed by an exponential cutoff,
so we take
\EQ
E_{\rm K}(k)={E_0\,(k/k_{\rm f})^4\over1+(k/k_{\rm f})^{17/3}}\,
\exp(-k/k_{\rm d}).
\EN
We use parameters representative of the simulations of B2000, so
$E_0=0.01$, $k_{\rm f}=5$ and $k_{\rm d}=25$. Like in B2000 we assume the
turbulence fully helical, so $H_{\rm K}=2kE_{\rm K}$ (e.g.\ Moffatt 1978).
The correlation functions $T(r)$ and $C(r)$ are then obtained via
\EQ
T(r)={2\over\tau}\int_0^\infty E(k)\,{j_1(kr)\over kr}\,\dd k
\equiv {\cal I}(E(k)),
\EN
and $C(r)= {\cal I}(F(k))/4$, 
where $j_1(x)=(\sin x-x\cos x)/x^2$ and $\tau$ is the correlation
time. (We use $\tau=4$, representative of the kinematic stage of Run~3
of B2000.)

\begin{figure}[t]\epsfbox{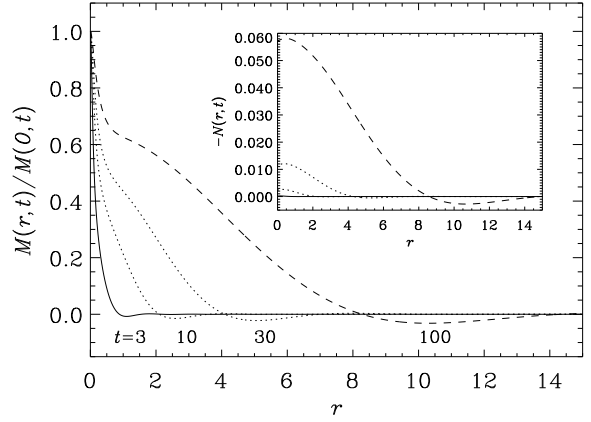}\caption[]{
Evolution of magnetic correlation functions for different times.
The correlation function of the magnetic helicity is shown in the inset.
$\eta=10^{-3}$.
}\label{Fpcor}\end{figure}

\begin{figure}[t]\epsfbox{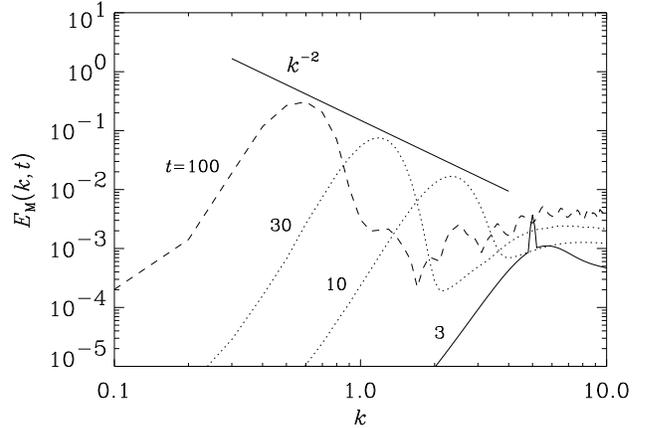}\caption[]{
Evolution of magnetic energy spectra.
Note the propagation of magnetic helicity and energy to progressively
larger scales. The $k^{-2}$ slope is given for orientation.
}\label{Fpspec}\end{figure}

We solve \Eqs{closure1}{closure2} using second order finite differences
and a third order time step on a uniform mesh in $0<x<L$ with up to
10,000 meshpoints and $L=10\pi$, which is large enough so that the
outer boundary does not matter.
In the absence of helicity, $C=0$, and without nonlinearity, $a=0$,
we recover the model of Novikov \ea (1983). The critical magnetic Reynolds
number based on the forcing scale is around 60. In the presence of helicity
this critical Reynolds number decreases, confirming the general result
that helicity promotes dynamo action (cf. Kim \& Hughes 1997, S99). In
the presence of nonlinearity
the exponential growth of the magnetic field terminates when the
magnetic energy becomes large. After that point the magnetic energy continues
however to increase nearly linearly. Unlike the case of the periodic box
(\Sec{Speribox}) the magnetic field can here extend to larger and larger
scales; see \Fig{Fpcor}. The corresponding magnetic energy spectra,
\EQ
E_{\rm M}(k,t)={1\over\pi}\int_0^L M(r,t)\,(kr)^3\,j_1(kr)\,\dd k,
\EN
are shown in \Fig{Fpspec}.

\epsfxsize=8.8cm

\epsfxsize=8.8cm

The resulting magnetic field is strongly helical and the magnetic
helicity spectra (not shown) satisfy $H_{\rm M}\la(2/k)E_{\rm M}$. The
development of a helicity wave travelling towards smaller and smaller 
$k$, as seen in \Fig{Fpspec}, is in agreement with the closure model 
of Pouquet \ea (1976). In the following we shall address the question 
of whether or not the growth of this large scale field depends on 
the magnetic Reynolds number (as in B2000). We have checked that to a 
very good approximation the wavenumber of the peak is given by
\EQ
k_{\rm peak}(t)\approx\alpha_\infty(t)/\eta_\infty(t).
\EN
This result is familiar from mean-field dynamo theory (see also S99)
and is consistent with simulations (B2000, section 3.5). Note that here
$k_{\rm peak}$ decreases with time because $\alpha_\infty$ tends to a
finite limit and $\eta_\infty$ increases. (This is not the case in the
box calculations where $k_{\rm peak}\ge2\pi/L$.)

\section{Resistively limited growth on large scales}

In an unbounded system the magnetic helicity,
$\bra{\AAA\cdot\BB}=6N(0,t)$, can only change if there is
microscopic magnetic diffusion and finite current helicity,
$\bra{\JJ\cdot\BB}=6H(0,t)$,
\EQ
\dd\bra{\AAA\cdot\BB}/\dd t=-2\eta\bra{\JJ\cdot\BB}.
\EN
The closure model of S97 and S99 also satisfies this constraint. (Note
that ambipolar and/or turbulent diffusion do not enter!) As explained in
B2000, this constraint limits the speed at which the large scale field
can grow, but not its final amplitude. One way to relax this constraint
is if there is a flux of helicity through open boundaries (Blackman \&
Field 2000, Kleeorin \ea 2000), which may be important in astrophysical
bodies with boundaries. Here, however, we consider an infinite system.

\begin{figure}[t]\epsfbox{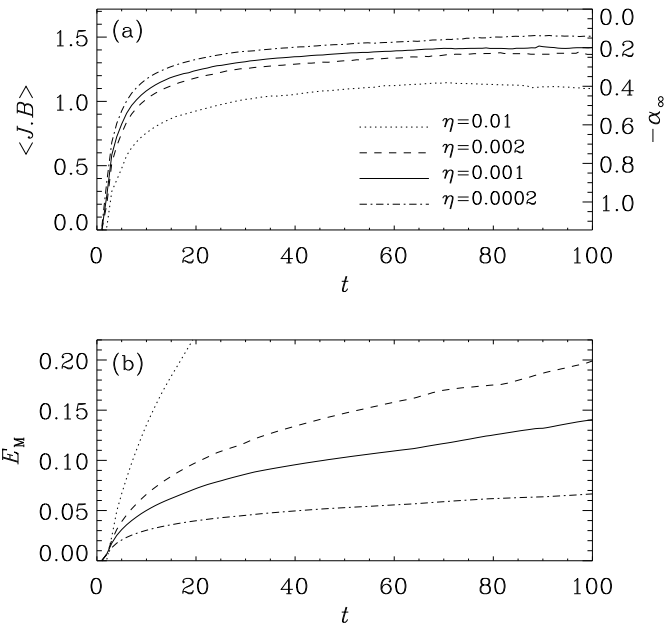}\caption[]{
(a) Evolution of $\bra{\JJ\cdot\BB}$ for different values of $\eta$.
The corresponding value of $\alpha_\infty$ is shown on the right hand
side of the plot. (b) The evolution of magnetic energy for the same
values of $\eta$.
}\label{Fpml}\end{figure}

In \Fig{Fpml} we show that, after some time $t=t_{\rm s}$,
$\bra{\JJ\cdot\BB}$ reaches a finite value. This value increases somewhat
as $\eta$ is decreased. In all cases, however, $\bra{\JJ\cdot\BB}$
stays below $\bra{\oo\cdot\uu}(2\tau/a)$, so that $|\alpha_\infty|$
remains finite; see \eq{asuppress}. A constant $\bra{\JJ\cdot\BB}$
implies that $\bra{\AAA\cdot\BB}$ grows linearly at a rate proportional
to $\eta$. However, since the large scale field is helical, and since
most of the magnetic energy is by now (after $t=t_{\rm s}$) in the large
scales, the magnetic energy is proportional to $\bra{\BB^2}\approx k_{\rm
peak}\bra{\AAA\cdot\BB}$, and can therefore only continue to grow at a
resistively limited rate, see \Fig{Fpml}.

\epsfxsize=8.8cm

\section{Conclusions}

Our results have shown that ambipolar diffusion (AD) provides a useful model
for nonlinearity, enabling analytic (or semi-analytic)
progress to be made in understanding nonlinear
dynamos. There are two key features that are shared both by this model and
by the full MHD equations: (i) large scale fields are the result of a nonlocal
inverse cascade as described by the $\alpha$-effect, and (ii) after some
initial saturation phase the large scale field
continues to grow at a rate limited
by magnetic diffusion. We reiterate that in astrophysical bodies the presence
of open boundaries may relax the helicity constraint. Furthermore, the
presence of large scale shear or differential rotation provides a means of
amplifying toroidal magnetic fields quite independently
of magnetic helicity, but this still requires poloidal fields for which
the above conclusions hold.

\begin{acknowledgements}
KS thanks Nordita for hospitality during the course of this work.
Use of the PPARC supported supercomputers in St Andrews and Leicester
is acknowledged.
\end{acknowledgements}

\end{document}